\documentstyle[12pt]{article}
\newcommand{\beq}{\begin{equation}\label}
\newcommand{\eeq}{\end{equation}}
\newcommand{\p}{\partial}

\newfont{\testb}{msbm10}
\begin{document} 

\title{The topology of multi-coupling deformations of CFT}
\author{Ulf Lindstr\"om\thanks{ul@physto.se} \hspace{0.5cm} Maxim Zabzine\thanks{On leave
from the Department of Theoretical Physics, St Petersburg University, Russia}\,\,\,\thanks{zabzin@vanosf.physto.se}\\
\textit{Institute of Theoretical Physics}\\
\textit{University of Stockholm}\\
\textit{Box 6730}\\
\textit{S-113 85  Stockholm, Sweden}}

\maketitle
.\vskip -12.0cm
\hfill USITP-97-08\\
\vskip 12.0cm
.\vskip -13.0cm
\hfill May, 1997\\
\vskip 13.0cm
.\vskip -14.0cm
\hfill hep-th/9705201
\vskip 14.0cm

Abstract: We discuss the topological properties of the manifold of
coupling constants for multi-coupling deformations of conformal field
theories. We calculate the Euler and Betti numbers and briefly discuss physical 
applications of these results.
\newpage

\section{Introduction}

The study of the  space of quantum field theories is an interesting an important
topic. 
In higher space-time dimensions not much is known
about this space, (although the recent results of Seiberg and
collaborators may be changing this \cite{sei}). In $1+1$ dimensions the
situation is much better, though, and all our considerations will concern
$D=2$. 

Recently geometrical methods have been applied to understanding the
renormalization group (RG) flow 
\cite{geo}. In these articles the RG flow was studied within the framework of
local differential geometry. Questions concerning the global topological
structure was not discussed. We are going to consider the global topological
structure of the RG flow from an ultraviolet to an infrared fixed point and the
relation between the topological invariants of the RG
flow and the local geometrical structure. Similar ideas were previously persued
by Vafa
\cite{Vafa}, but whereas he considers the space of {\em all} 2D quantum field
theories (QFTs), our
considerations are restricted to the space of coupling constants relevant for
the description of the flow of a QFT from its ultraviolet (UV) to its infrared
(IR) fixed point.

The fixed points of the RG flow are conformal
field theories (CFTs). Let us consider a two-dimensional QFT with one coupling
constant. It can be viewed  as a deformation of a CFT by a spinless operator
$\Phi$:
\beq{a1}
S = S_{CFT} + \lambda \int d^{2} x \Phi(x).
\eeq
This CFT describes the UV limit of the theory if the
scaling dimension of $\Phi$ is $d < 2$, so that the perturbation is by
a relevant operator, and $\lambda$ has mass dimension $2-d$. If $d=2$
the QFT is asymptotically free and $\Phi$ is a marginal operator. The
UV CFT describes the short distance behavior of an off-critical
QFT. The long distance behaviour is
described by an IR CFT. When the coupling constant is
dimensionful, (\ref{a1}) can be considered effectively as a
perturbation away from an IR CFT by an irrelevant operator. 

The situation described above is the general one for $1+1$
off-critical QFTs. Particular deformations may result in integrable
systems. In that case one can solve the theory non-perturbatively
within a bootstrap program \cite{muss}.

The really interesting problem is multi-coupling deformations of a
CFT. However, integrability is lost, e.g., already in a two-coupling
deformation of the Ising model (or the tricritical Ising
model) \cite{muss}. This may be traced back to the different
null-vector conditions satisfied by the perturbing relevant fields. It
is an important task to find tools to study nonperturbative effects in
such models. The aim of the present letter is to provide a step in
this direction.

The content of the letter is as follows: In section \ref{1} we outline
the geometric interpretation of manifold of coupling constants. In
section \ref{2} we show that Zamolodchikov's C-function  can be
considered as a Morse function for this manifold. We also calculate
the Euler and Betti numbers. The topology of a manifold corresponding
to deformations with even number of coupling constants  differs from
that corresponding to an odd number of deformations. In section
\ref{3} we briefly discuss some physical applications of our knowledge
of the topological properties of the  manifold of coupling constants. The 
concluding remarks are summarized in section \ref{4}.  
  
\section{The manifold of coupling constants}\label{1}

We study the deformation of a CFT by some finite number of relevant
operators. Near the fixed point we have,
\beq{b1}
S = S_{CFT} + \sum_{i=1}^{N} \lambda_{i} \int d^{2} x \Phi_{i} (x).
\eeq
The operators $\Phi_{i}$ have dimensions $d_{i}$ with respect to the UV CFT. The 
dimensionful coupling constants can be represented as 
\beq{b2}
\lambda_{i} = g_{i} \mu^{2 - d_{i}},
\eeq
where $\mu$ is a mass parameter and the $g_{i}$'s are dimensionless
coupling constants. We will consider the action (\ref{b1}) to describe
the RG flow from an UV CFT to an IR CFT for a QFT. 
Basic examples of this situation are provided by the deformations
of the unitary minimal models with central charges \cite{muss}
\beq{11}
c = 1 - \frac{6}{m (m+1)},\,\,\,\,\,\,\,\,m > 1.
\eeq
The number of relevant operators for these models is $2(m - 2)$.  

The appropriate objects in a renormalization group setting are in fact
the
running coupling constants $g_{i}(\mu)$ which satisfy the equations
\beq{b3}
\mu \frac{d g_{i}}{d \mu} = \beta_{i} (g)
\eeq
$$g_{i}(\mu_{*}) = g_{i0}$$
where $g_{i0}$ are defined at some
renormalization scale $\mu_{*}$, and $\beta_{i}(g)$ are
the Gell-Mann-Low $\beta$ functions. We define the manifold of coupling 
constants $\cal M$
 to be the manifold
coordinatized by $\{g_{i}\}$. The geometry of this manifold will be
defined by the renormalization of the QFT. The RG is one-parameter
group of motions on $\cal M$ defined through (\ref{b3}).

The importance of the geometry of $\cal M$ is clear from the fact that,
introducing a renormalization scale $\mu$, one
can treat the correlation functions as functions on $\cal
M$. Namely, consider a correlation function 
\beq{ff}
D(p_{1}, p_{2}, ..., p_{n}, g_{1}, g_{2}, ..., g_{N}) = <O_{1}(p_{1}) 
O_{2}(p_{2}) ... O_{n}(p_{n})>,
\eeq
where $p_{1}, p_{2}, ..., p_{n}$ are the momenta and $O_{1}, O_{2},
..., O_{n}$ some local operators. As a consequence of the RG equations
this function can be rewritten as
\beq{b4}
D(p_{1}, p_{2}, ..., p_{n}, g_{1}, g_{2}, ..., g_{N}) = \mu^{d} F 
(\frac{p_{1}}{\mu}, \frac{p_{2}}{\mu}, ..., \frac{p_{n}}{\mu}, g_{1}, g_{2}, 
..., g_{N}),
\eeq
where $d$ is the full mass dimension of the correlation function. This is a
consequence of the RG equations. If we fix $p_{1}, p_{2}, ..., p_{n}$,
$F$ becomes a function on $\cal M$. This is equivalent to introducing
the normalization scale $\mu$ in the correlation function 
\beq{b5}
<O_{1}\, O_{2}\, ...\, O_{n}>|_{p_{1}=\mu, p_{2}=\mu, ..., p_{n}=\mu}
\eeq 
In this manner we may consider the correlation functions as functions
on $\cal M$.

\section{The C-theorem and the topology of $\cal M$}\label{2} 

Two dimensional QFT offers a unique opportunity to study the global
topology of $\cal M$ using Zamolodchikov's C-theorem \cite{Zam}. This
theorem about the RG flow for $1+1$ QFT
establishes that there exists a function $C(g_{1}, g_{2}, ..., g_{N})$
defined on $\cal M$ which is non-increasing along the RG
trajectories and is stationary at the fixed points only.There it
coincides with the central charge of the corresponding CFT. The proof
of 
the C-theorem assumes renormalizability, rotational and translational 
invariance, reflection positivity and conservation of the 
stress-energy tensor.

The first observation on the relevance of the C-theorem to the
topology of 2D QFTs was made by Vafa \cite{Vafa} in discussing the
space of {\em all} $2D$ QFTs. We will treat an arbitrary QFT with CFT
fixed points and consider the C-function as a
Morse function on the relevant $\cal M$. Note that this restricts us to the
RG flow between the UV and the IR fixed points. What happens after that
(branching et c) concerns other QFTs. In Morse theory knowledge of the
critical points of a function on a manifold and the behavior of the function
near  these critical point carries information about the global topological 
propeties of manifold. The C-function on $\cal M$ corresponding to the  QFT
with action (\ref{b1}) has two critical points, ( the UV and IR  points), and at
these points it satisfies the constraint
\beq{c1}
\frac{\p C}{\p g_{i}} = 0, \,\,\,\,\,\, i = 1, 2, ..., N.
\eeq
In a neighbourhood of the UV critical point the C-function can be 
represented as
\beq{c2}
C(g_{1}, g_{2}, ..., g_{N}) = c - \sum_{i=1}^{N} 3 (2 - d_{i}) g_{i}^{2} + O 
(g^{3})
\eeq
where c is the central charge of the UV CFT. The Hessian for the C-function
is thus
\beq{c3}
det (\frac{\p^{2} C}{\p g_{i} \p g_{j}}) = (-1)^{N} \prod_{i=1}^{N} 6(2 - d_{i}) 
+ O (g).
\eeq
The symmetric bilinear form in neighbourhood of critical point is
\beq{c4}
dC^{2} = \sum_{i=1}^{N} -6(2 - d_{i}) d g_{i}^{2}.
\eeq
From (\ref{c3}) and (\ref{c4}) we see that the UV critical point is 
nondegenarate and that the index\footnote{The index of the critical
point is number of negative squares in form $d^{2}C$.} of this point
is equal to $N$. 

The IR critical point has no relevant directions (cf the comment below
(\ref{a1})). This means that the index of the IR point is $0$. 
The nondegeneracy of the IR point is not obvious. We believe
that 
the theory near the IR point can be effectivly described through a
perturbation of the IR CFT by irrelevant operators. 
The C-function may then be represented also in neighbourhood of the IR
point 
as
in (\ref{c2}) and we may conclude that this point too is nondegernerate.

Let us make some mathematical comments. For our considerations below to be
valid, the manifold $\cal M$ and the function C must satisfy certain
restrictions. The manifold $\cal M$ must be smooth and separable. This
is true in the present case, since $\cal M$ has a Riemannian
structure given by Zamolodchikov's metric. The function C has to be
bounded from below, (in our case $C>0$), have a finite number of
(nondegenerate) critical points and near every critical point there
should exist an $\epsilon > 0$ such that the set $\{x\in \cal
M:$ $c-\epsilon\leq C(x) \leq c+\epsilon\}$ is compact. Here $c$ is the value of C
at the critical point \cite{post}. This last requirement 
can be realized on $\cal M$ by a suitable coordinatization near
the critical points.

We have shown that the C-function is a Morse function with two 
nondegenarate critical points. Let us now calculate the Euler number
for $\cal M$,
\beq{c5}
\chi ({\cal M}) = (-1)^{N} + (-1)^{0} = \left \{
                                        \begin{array}{ll}
                                           0, & N\,\,\, odd\\
                                           2, & N\,\,\, even.
                                        \end{array}
                                  \right.  
\eeq
A relevant object for further discussing the topology of $\cal M$ is
the following polynomial corresponding to the C-function \cite{fom}
\beq{c6}
K(y) = y^{N} + 1.
\eeq
With every manifold one also associates a Poincar\'e polynomial defined by
\beq{c7}
P(y) = \sum_{k=0}^{N} dim H^{k}(M)\, y^{k},
\eeq
where $H^{k}(M)$ are the homology groups. Morse theory tells us
that in 
general these two polynomials connected the following way
\beq{c8}
K(y) - P(y) = (1 + y) T(y),
\eeq
where $T(y)$ is polynomial with positive and integer coefficients. For
$\cal M$ we obtain the dimensions of the homology groups ($dim
H^{k}(M)$), called the Betti numbers
\beq{c9}
dim H^{0}(M)=1,\,\,\,\, dim H^{1}(M) = 0,\,\, ...,\,\, dim H^{N-1}(M) = 
0,\,\,\,\, dim H^{N}(M) = 1.
\eeq
We have found both the Euler number and the Betti numbers for
$\cal M$ from the C-function. For two dimensional QFT we thus have the
exceptional situation that one can obtain
information about the global RG topology from general principles.

In many interesting cases of multi-coupling deformation of CFT the 
manifold $\cal M$ can be consider to be a compact, closed and
connected. 
This is possible, e.g., when theory has an IR fixed point at finite
values 
of the coupling constants. Also an IR point at infinite values of 
the coupling constants is included in considering $\cal M$\footnote{In
analogy to viewing $R^{2} \times \{ \infty \}$ as homeomorphic to the
sphere $S^{2}$}, because we can think of this manifold as compact. It
is a well-known fact \cite{fom} that if there is a smooth function C 
with only two critical points (perhaps degenerate) then this manifold
is 
homeomorphic to the sphere $S^{n}$.

\section{Physical applications}\label{3}

Global topological invariants may be expressible in terms of local geometric 
quantities. A well known example is the Gauss-Bonnet theorem for two
dimensional compact manifolds. We will briefly discuss an application
of 
this theorem. Let us consider a two-coupling deformation of an UV CFT
\beq{f1}
S = S_{CFT} + g_{1} \mu^{2-d_{1}} \int d^{2}x \Phi_{1} (x) + g_{2} \mu^{2-d_{2}} 
\int d^{2}x \Phi_{2} (x).
\eeq
We introduce the function (cf. (\ref{b5}),
\beq{f2}
G_{ij}(g_{1}, g_{2}) = x^{4} <\Phi_{i}(x) \Phi_{j}(0)> |_{x^{2}=x_{0}^{2}}
\eeq
where $x_{0}$ is the renormalization scale. The symmetric matrix
$G_{ij}(g_{1}, g_{2})$ is positive definite and may be thought of as
a metric on $\cal M$ \cite{Zam}. For this metric we define the
curvature form $R(g_{1}, g_{2})$. The Gauss-Bonnet theorem gives us
a relation between the Euler number and the integral of the curvature form
\beq{f3}
\chi({\cal M}) = \frac{1}{2\pi} \int_{\cal M} R = \frac{1}{2\pi}
\int_{\cal M} K dS.
\eeq
Here $K$ is the Gaussian (extrinsic) curvature and $dS$ is the
area element. For the case of a two-coupling deformation we know the
Euler number and metric (\ref{f2}). One may then write the curvature
form 
in terms of the metric and use the Gauss-Bonnet theorem to obtain
\beq{f4}
4\pi = \int_{\cal M} S (x^4<\Phi_{i}(x) \Phi_{j}(0)>|_{x^{2}=x_{0}^{2}}) dg_{1} 
\wedge dg_{2},
\eeq
where S is function defined by
\beq{f5}
S(G_{ij}) dg_{1} \wedge dg_{2} = R = K dS.
\eeq
We thus obtain the non-perturbative integral equality (\ref{f4}) for
the correlation function $<\Phi_{i}(x) \Phi_{j}(0)>$. We believe that
this equality can be used to study non-perturbative effects in
multicoupling deformations of 
CFTs.  The relation (\ref{f4}) bears a similarity to Cardy's sum
rule \cite{cdy}
\beq{cd}
c_{UV}-c_{IR}={\frac{3}{4\pi}}\int_{|x|<\epsilon}d^2x x^2<\Theta (x)\Theta (0)>,
\eeq
where $c_{UV}$ and $c_{IR}$ are the central charges of the UV and IR conformal
theories and $\Theta$ is the trace of the stress-energy tensor, which
may be expressed in terms of the relevant operators $\Phi_{i}(x)$ as
\beq{cart}
\Theta = 2\pi\beta_i(g)\Phi_{i}(x).
\eeq
(We hope to return to the exact relation in the future).

The above discussion is immediately generalised to an arbitrary even number
of 
coupling constants.

 We introduce the Euler class $e({\cal M})$ which
is 
an element of the cohomology group $H_{N}({\cal M}, Z)$. The Euler
class is expressible in terms of the curvature form. Again, we can
start from the metric $x^4<\Phi_{i}(x)\Phi_{j}(0)>|_{x^{2}=x_{0}^{2}}$ and
construct different geometric structures. For example, the integral of the Euler
class over $\cal M$ gives 
\beq{f6}
\chi ({\cal M}) = \int_{\cal M} e({\cal M}).
\eeq
where the Euler number is two. We thus again find a nontrivial equality for
the correlation function.  

\section{Conclusions}\label{4}

We have shown that in $1+1$ QFT the C-function has the properties of a
Morse function on the manifold coupling constats $\cal M$. If the
normalization point is introduced in the correlation functions, we can
treat them as functions defined on $\cal M$. The Zamolodchikov
metric on $\cal M$ gives it a Riemannian structure. The
Gauss-Bonnet-Chern-Avez theory relates the global topological
invariant (the Euler number) to the differential-geometric structure
(the Euler class) on an even dimensional manifold. Using this theorem
we obtained a nonperturbative equality for correlation functions for 
multicoupling deformations of a CFT.

\section{Acknowledgements}

We are grateful to F. Bastianelli, J. Grundberg and G. Shore for constructive 
criticism.
M.Z. thanks ITP, University of Stockholm for hospitality. M.Z. was
supported by a grant of the Royal Swedish Academy of
Sciences. U.L. acknowledges support from NFR under contract No F-AA/FU
04038-312 and from NorFA under contract No 9660003-0.

\end{document}